\documentclass[prd,showpacs,preprintnumbers,nofootinbib,aps]{revtex4}
\usepackage{subeqnarray}
\usepackage{epsfig,amsmath,amssymb}
\usepackage{mathrsfs}
\usepackage[usenames,dvipsnames]{color}
\usepackage[pagebackref=true, colorlinks=true]{hyperref}
\definecolor{redish}{rgb}{0.7,0.2,0.0}  
\definecolor{bluish}{rgb}{0.2,0.5,0.8}
\definecolor{MyLightMagenta}{cmyk}{0.3,0.7,0.3,0.3}
\hypersetup{linkcolor=redish,          
                  citecolor=blue,        
                  filecolor=magenta,      
                  urlcolor=bluish}          
\DeclareFontFamily{U}{rsfs}{}         
\DeclareFontShape{U}{rsfs}{m}{n}{<5> rsfs5 <6><7> rsfs7          %
  <8><9><10><10.95><12><14.4><17.28><20.74><24.88> rsfs10}{}     %
\DeclareMathAlphabet{\mathfs}{U}{rsfs}{m}{n}                     %

\newcommand{\ba}{\nopagebreak[3]\begin{eqnarray}}
\newcommand{\ea}{\end{eqnarray}}
\newcommand{\bii}{\begin{itemize}}
\newcommand{\eii}{\end{itemize}}
\newcommand{\nn}{\nonumber}

\newcommand{\f}{\frac}

\def \d{\delta}

\def \l{\ell}
\def \g{\gamma}
\def \e{\epsilon}
\def \lp{\l_p}
\def \j{\sqrt{j(j+1)}}

\def \lm{\lambda}

\def \s{\sigma}

\def \H{\mathcal{H}}
\def \P{\mathcal{P}}

\def \mic{{\it microcanonical}}
\def \sj{s_j^{\star}}
\def \H{\mathcal{H}}
\def \P{\mathcal{P}}

\def \th{\theta}
\def \tn{\theta_0}
\def \3{\sqrt 3}
\begin{document}

\title{`Quantum Hairs' and Entropy of Quantum Isolated Horizon from Chern-Simons Theory}
\author{Abhishek Majhi}%
 \email{abhishek.majhi@saha.ac.in}
\affiliation{Saha Institute of Nuclear Physics, Kolkata 700064, India}%
\author{Parthasarathi Majumdar}
\email{bhpartha@gmail.com}
\affiliation{Ramakrishna Mission Vivekananda University, Belur Math 711202, India}%

\begin{abstract}

We articulate the fact that the loop quantum gravity description of the quantum macrostates of black hole horizons, modeled as Quantum Isolated Horizons (QIHs), is completely characterized in terms of two independent integer-valued `quantum hairs', viz,. the coupling constant $(k)$ of the quantum $SU(2)$ Chern-Simons theory describing QIH dynamics, and the number of punctures $(N)$ produced by the bulk spin network edges piercing the isolated horizon (which act as pointlike sources for the Chern- Simons fields). We demonstrate that the microcanonical entropy of macroscopic (both parameters assuming very large values) QIHs can be obtained directly from the microstates of this Chern-Simons theory, using standard statistical mechanical methods, without having to additionally postulate the horizon as an ideal gas of punctures, or incorporate any additional classical or semi-classical input from general relativity vis-a-vis the functional dependence of the IH mass on its area, or indeed, without having to restrict to any special class of spins. Requiring the validity of the Bekenstein-Hawking area law relates these two parameters (as an equilibrium `equation of state') and  consequently allows the Barbero-Immirzi parameter to take any real and positive value depending on the value of $k/N$. The logarithmic correction to the area law obtained a decade ago by R. Kaul and one of us (P.M.), ensues straightforwardly, with precisely the coefficient -3/2, making it a signature of the loop quantum gravity approach to black hole entropy.
\end{abstract}

\maketitle
\section{Introduction}

Perhaps the clearest description of the quantum states of generic (extremal or non-extremal) four dimensional black hole horizons in the presence of matter and/or radiation (which do not cross the horizons) is in terms of the quantization of the classical {\it isolated horizon} ({\bf IH}) phase space which yields the kinematical Hilbert space for the {\it quantum isolated horizon} ({\bf QIH}) \cite{qg1,qg2} derived from {\it loop quantum gravity} ({\bf LQG}). In this description,  at the classical level, the IH is considered as an inner boundary of spacetime with boundary conditions imposed upon it which (a) do not require that the ambient spacetime be stationary and (b) hold the area of spatial foliations (time-slice) of the IH fixed, precluding matter or radiation from crossing it. It has been shown in \cite{abf} that originating from these boundary conditions is a Hamiltonian structure of the degrees of freedom of a generic IH and their dynamics, in terms of an $SU(2)$ {\it Chern-Simons} ({\bf CS}) theory, where the CS connection (on the chosen time-slice) belongs to a one-parameter family of linear combination of components of the usual Levi-Civita connection of general relativity, pulled back to the spatial slice. The CS connection couples to the bulk spacetime geometry, namely the bulk triads on the spatial slice of bulk spacetime, which act as sources for the CS connection, with the CS coupling being proportional to the classical area of the spatial slice of the IH. A somewhat more direct approach to the problem, albeit restricted to a static spacetime, namely, the Schwarzschild spacetime, has been given in \cite{km11} where the same geometric variables as used in \cite{abf} are evaluated explicitly, with the manifest emergence of the $SU(2)$ CS description of horizon degrees of freedom and their dynamics, namely their coupling to bulk geometry. Classically therefore, a black hole horizon is nothing but an $SU(2)$ CS theory of connection fields on an IH (a generalization, to reiterate, of an event horizon to non-stationary ambient spacetimes in the bulk) which are minimally coupled to sources consisting of bulk spacetime tetrads. Since tetrads are formally like the `square root' of the metric, at the classical level, the relationship to the spacetime metric is clear. 

 In the quantum description, the physical states of a QIH are the gauge singlet states of an $SU(2)$ Chern Simons (CS) theory coupled to the punctures endowed with $SU(2)$ spins deposited on the QIH by the intersecting edges of the bulk spin network describing the bulk quantum geometry. In other words, the QIH is taken to have the topology of a two-sphere with punctures carrying spins induced by the floating lattice known as the spin network whose edges carry spins. Such configurations are expected to arise as solutions of the full quantum dynamical equations describing bulk quantum geometry, in particular the quantum Hamiltonian constraint. Whether or not this actually happens remains a question for the future. In this situation, it is worthwhile to investigate properties of the kinematical Hilbert space to see how the issue of black hole entropy can be addressed.    

There are a plethora of approaches to black hole entropy, many semiclassical ones with diverse claims for the black hole entropy having logarithmic corrections to the area law \cite{sol,fursaev,sen1,sen2,sen3,sen4}, with various coefficients -- some positive and some negative. In all these computations, one usually computes the entropy of quantum `matter' fields (including gravitons) coupled to the classical background spacetime metric of a black hole. Even in approaches where the classical metric is `integrated over', the functional integral over metrics is invariably saturated by the classical black hole metric. In other words, non-perturbatively large quantum fluctuations of the spacetime - which cannot meaningfully be separated into a classical background metric and its fluctuations - are usually ignored in these computations. Thus, these computations do {\it not} take into account the entirety of quantum states corresponding to quantum spacetime fluctuations of a black hole spacetime. Rather, they account for non-gravitational states {\it entangled} with the horizon, as discussed in ref. \cite{bomb,sred}. In contrast, the LQG approach focuses on the quantum states describing the {\it quantum} geometry of the horizon, without assuming any entanglement with quantum matter states in its vicinity. The classical metric of the black hole plays no role in this approach. The LQG computation thus yields what one may call the {\it gravitational} (or `spacetime') entropy of a black hole, as opposed to the non-gravitational entanglement entropy computed in other approaches, which depends rather directly on the classical black hole metric in all cases. Indeed, the two together would most likely constitute the total entropy of a quantum black hole spacetime, so that the results of both directions of computation are complimentary rather than competitively comparable. In other words, the entanglement entropy considered in the other approaches must be an additional contribution to the horizon entropy, over and above the spacetime entropy computed within the LQG approach.             

The description of a QIH in terms of CS states coupled to bulk quantum geometry is itself a radical departure from the general relativistic description of a classical event horizon. Of course, a QIH does not radiate or accrete matter/radiation, and can only be a part of a more complete description of a radiant black hole comprising of Trapping \cite{hay} or Dynamical \cite{ashkris} horizons. Recently, such horizons have been argued to emit Hawking radiation \cite{pranz} with the standard black body spectrum, raising hopes that a deeper understanding of black holes within the LQG approach might be around the corner. However, there are semi-classical assumptions that appear to be necessary to supplement the premises of LQG, which manifest a lingering incompleteness in basic understanding of the quantum geometry of IHs. 

The object of the present paper is to articulate a fact that has already been implicit in recent literature \cite{gp}, however, {\it without} the extra semi-classically motivated assumptions used earlier : that a complete description of  the macrostates of a QIH is possible with two integer-valued quantum parameters (`quantum hairs'), namely the CS coupling constant $k$ related to the classical horizon area $A_{cl}$, and the number $N$ of punctures to which the CS fields couple. Indeed, the microcanonical entropy can be obtained directly from the formula derived fourteen years ago by one of us (P. M.) together with R. K. Kaul \cite{km98} for the total number of $SU(2)$ singlet states coupled to a fixed set of spins $j_1, j_2, ...,j_N~,~j_i \in [1/2,k/2] \forall i=1,...N$. The independence of these two parameters $k$ and $N$ has been quite apparent at the formulation stage, while deriving this formula, even though their physical role and the precise path whereby a relation between the two emerges only becomes obvious through ref. \cite{gp}, (albeit upon their invoking semiclassical arguments which we have made no use of).  Summing this formula over all spins $\{j_i\}$ and using the Multinomial Expansion of elementary algebra, this degeneracy can be expressed in terms of sums over spin {\it configurations}, i.e., number of punctures $s_j \in[0,N]$ for spins $j=1/2,...,k/2$.  This upper bound on allowed spin-values follows directly from the unitarity of the two dimensional conformal field theory (viz., the $SU(2)_k$ WZW model) living on the boundary $S^2$, whose conformal blocks capture the degeneracy information of the CS states on the QIH, as stated in section(\ref{sec2}) below.

Using standard tenets of equilibrium statistical mechanics, one then looks for the `Most Probable Spin Configuration' by maximizing the microcanonical entropy subject to the constraints \cite{pathria} of a fixed $N$ and fixed large QIH area differing from the classical area only to $O(\lp^2)$, where $\lp$ is the Planck length. This extremization procedure leads to an equilibrium `equation of state' relating the two quantum parameters $k$ and $N$ describing the macrostates. Requiring that the resulting formula for the microcanonical entropy of a macroscopic ($k \ggg 1,~N \ggg 1$) QIH reproduce the Bekenstein-Hawking area law in terms of classical IH area $A_{cl}$ to leading order, alongside a term proportional to the quantum hair $N$, imposes a restriction on the Barbero-Immirzi parameter $\gamma$ in the definition of $k \equiv A_{cl} / 4\pi \gamma \lp^2$, that this must lie within a specific interval on the real line \cite{majhi1}. Subleading corrections to the area law ensue naturally from the derived formula, especially the leading logarithmic (in {\it classical} area) correction with coefficient $-3/2$ found longer than a decade ago by Kaul and Majumdar \cite{km00} for a dominant class of spins, as we shall see in the sequel. These results have also been rederived recently in ref. \cite{per1,per2,per3,per4}

 It may be asked how precisely this paper builds upon from extant literature on microcanonical entropy by counting of CS states in LQG. We wish to note here that in those earlier papers, the formulation of QIH states in terms of two independent parameters (the so-called `quantum hair' as espoused in \cite{gp}) was not used explicitly. In some papers, the entropy is computed simply by counting states of an ideal gas of punctures, so that the CS underpinning of the QIH states is not used explicitly. In some of the very early LQG literature, the computation does not quite use the connection with two dimensional $SU(2)_k$ WZW models which is crucial to our approach here. Still in some other papers, the counting is restricted to a `dominant' spin configuration on the punctures, so that even though the approximation is not incorrect, the approach appears to lack generality.
 
The important value-addition  in this paper is to provide a direct link of the `quantum hair' scenario \cite{gp} to the LQG formulation of QIH  macrostates in terms of $SU(2)$ CS states coupled to punctures carrying spin, without any trappings of a semi-classical nature (like \cite{gp}) depending on classical metrical properties, nor any restriction to a specific class of spin values at the punctures. This connection of the CS formulation of the QIH to the conceptual understanding gained recently in ref. \cite{gp} of macrostates in terms of two integer-valued parameters $k$ and $N$, leading eventually to the same microcanonical entropy as found earlier \cite{km00} thus ties up the older formulations with contemporary ideas in this field on a unified footing. While many of the extant papers in the recent literature on QIH entropy within the LQG approach simply count states of an ideal gas of spins, without much allusion to the underlying theory of QIHs in terms of quantum CS states, our approach here has been to underline the CS theory underpinning, and to make the argument as self-contained as possible within the original LQG approach pioneered by Ashtekar and coworkers \cite{dole,mei}, while concomitantly relating to more recent aspects of the literature. This paper has some overlap with the recent review of Kaul \cite{kaul12} as far as some parts of the calculations are concerned, but the emphasis on direct link-up with the CS theory perspective and some of the details are different and are, hopefully, of inherent merit.    

\section{The Hilbert Space and Physical Chern-Simons states of the QIH}\label{sec2}

The Hilbert space of a quantum spacetime admitting QIH as an inner boundary is given by $\H=\H_V\otimes\H_S$ modulo gauge transformations, where $V$ denotes bulk and $S$ denotes boundary(QIH) at a particular time slice\cite{qg1,qg2}. Mathematically, if the 4d spacetime $({\bf R}\times\Sigma)$ admits a 3d IH $(\Delta)$ as null inner boundary, then $S\equiv \Delta\cap\Sigma$ denotes a cross-section of the IH \cite{ashkris}.  Hence, a generic quantum state of the spatial geometry of such a spacetime can be written as $|\Psi\rangle=|\Psi_V\rangle\otimes|\Psi_S\rangle$, where $|\Psi_V\rangle$ is the wave function\footnote{Strictly speaking, these are actually {\it functionals} of the $SU(2)$ spatial connection variables and a smooth {\it function} of generalized gauge-invariant connections, the holonomies along the edges of the oriented graph \cite{ashlew}, popularly known as the spin network.} corresponding to the volume$(V)$ or bulk states represented by an oriented graph, say $\Gamma$, consisting of edges and vertices \cite{ashlew} and $|\Psi_S\rangle$ denotes a generic quantum state of the QIH. $|\Psi_S\rangle \in\H_S\equiv$ the Hilbert space of the CS theory coupled to the punctures $\{\P\}$ made by the bulk spin network $\Gamma$ with the IH endowing them with the spin representations carried by the respective piercing edges which are solely responsible for all the relevant features of the QIH, the most important being the quantum area spectrum of the QIH. To be precise, for a given $N$ number of punctures, with spins $(j_1,\cdots,j_N)$, the QIH Hilbert space is given by $\H_S\equiv \text{Inv}(\otimes_{i=1}^N\H_{j_i})$ where `Inv' denotes the invariance under the local $SU(2)$ gauge transformations on the QIH. Now, as it is seen that at the quantum level the full Hilbert space is the direct product space of the bulk and boundary Hilbert spaces, a generic quantum state of the QIH (boundary) can be written in terms of basis states on $\H_S$, independent of the bulk wave function. Hence, one should understand that a basis state of the QIH Hilbert space is actually a generic quantum state of the full Hilbert space, since the bulk part of the wave function is a linear combination of the basis states of the bulk geometry. In other words, a given {\it spin configuration} on the QIH admit all possible graphs $(\Gamma$-s) in the bulk consistent with the given configuration. This spin configurations provide the area eigenstate basis, which is the all important material in the context of QIH entropy. Such a basis state of the QIH Hilbert space is denoted by the ket $|\{s_j\}\rangle$. This is an eigenstate of the area operator associated with the QIH, having the area eigenvalue given by $\hat A_S|\{s_j\}\rangle=8\pi\g\lp^2\sum_{j=1/2}^{k/2} s_j\j|\{s_j\}\rangle$. Such a spin configuration (eigenstate) has a $(N!/\prod_j s_j!)$-fold degeneracy due to the possible arrangement of the spins yielding the same area eigenvalue. Hence, a generic quantum state of the QIH can be written as 
\ba
|\Psi_S\rangle=\sum_{\left\{s_j\right\}}c[\left\{s_j\right\}]|\left\{s_j\right\}\rangle\nn
\ea 
where $|c[\left\{s_j\right\}]|^2=\omega[\left\{s_j\right\}]$(say) is the probability that the QIH is found in the state $|\left\{s_j\right\}\rangle$. Hence, a generic quantum state of the spacetime, admitting QIH as an inner boundary, may now be written as $|\Psi\rangle\equiv |\Psi_V\rangle \otimes \sum_{\left\{s_j\right\}}c[\left\{s_j\right\}]|\left\{s_j\right\}\rangle$. 

\par
The computation of the microcanonical entropy formula of Kaul and Majumdar \cite{km98} proceeds from the expression for the number of conformal blocks of $SU(2)_k$ WZW model on a 2-sphere with marked points (punctures) carrying spin which, in the seminal work of  Witten \cite{wit}, has been shown to give the dimensionality of the $SU(2)$ singlet part of the Hilbert space of CS states on ${\bf R} \times S^2$ coupled to punctures on the $S^2$. Using this remarkable connection, the fusion algebra and the Verlinde formula, the degeneracy of the microstates is expressed as \cite{km98}
\ba
\Omega(j_1, \cdots ,j_N)=\f{2}{k+2}\sum^{k+1}_{a=1}\f{\sin\f{a\pi(2j_1+1)}{k+2}\cdots\sin\f{a\pi(2j_N+1)}{k+2}}{\left(\sin\f{a\pi}{k+2}\right)^{N-2}}\label{cs}
\ea
which can be alternatively recast as a linear combination of Kronecker deltas \cite{km98}, explicitly manifesting the singlet nature of the physical states :
\ba
\Omega (j_1, \cdots, j_N)&=&\sum_{m_1=-j_1}^{j_1}\cdots\sum_{m_N=-j_N}^{j_N}\left[\d_{\left(\sum_{p=1}^{N}m_p\right),0} -\f{1}{2}~\d_{\left(\sum_{p=1}^{N}m_p\right),1}-\f{1}{2}~\d_{\left(\sum_{p=1}^{N}m_p\right),-1}\right]\nn
\ea
For the calculations we shall use the expression (\ref{cs}). To obtain the total number of conformal blocks, this expression must be summed over all possible spin values at each puncture.
\ba
\Omega(N,k)=\sum_{j_1,\cdots ,j_N}~\Omega(j_1, \cdots ,j_N)\label{ms}
\ea

Now, since we will apply the method of most probable distribution to find the microcanonical entropy of the QIH, it is convenient to recast eq.(\ref{ms}) as sum over spin-configurations, i.e., the number of punctures carrying a specific spin $j$, for all possible values of $j$, becomes the dynamical variable. 

\ba
\Omega(N,k)=\sum_{\left\{s_j\right\}}
~\Omega[\left\{s_j\right\}]\label{nk}
\ea
where 
\ba
\Omega[\left\{s_j\right\}]=&&\f{N!}{\prod_{j}s_j!}g[\{s_j\}]\label{ms1}
\ea
where $j$ runs from $1/2$ to $k/2$ as usual and
\ba
g[\{s_j\}]=\f{2}{k+2}\sum^{k+1}_{a=1}\sin^2\f{a\pi}{k+2}\prod_{j}\left\{\f{\sin\f{a\pi(2j+1)}{k+2}}{\sin\f{a\pi}{k+2}}\right\}^{s_j}\label{gsj}
\ea

The combinatorial factor  in eq.(\ref{ms1}) reflects the {\it statistical} distinguishability of the punctures, a property inherited by the punctures in the quantization procedure of the classical IH due to the nontrivial holonomies of the Chern-Simons connection on the IH along disjoint closed loops about the punctures\cite{qg2}.

\subsection{Multinomial Expansion : The Link}\label{metcl}

It is worth mentioning that, even though eq.(\ref{ms1}) resembles the formula for the microstates of an ideal gas of spins obeying Maxwell-Boltzmann statistics, it has been derived directly from the formula (\ref{cs}) within the present scenario of QIH.  One should keep in mind that eq.(\ref{ms1}) is just another form of eq.(\ref{ms}) being written in a different basis for convenience of the statistical formulation. Hence, for the sake of clarity, let us derive eq.(\ref{ms1}) directly from eq.(\ref{ms}).
Using eq.(\ref{cs}) one can write eq.(\ref{ms}) in the following explicit form
\ba
\Omega(N,k)&=&\f{2}{k+2}\sum^{k+1}_{a=1}\sin^2\f{a\pi}{k+2}\sum_{j_1,\cdots, j_N}\prod_{r=1}^N\left\{\f{\sin\f{(2j_r+1)a\pi}{k+2}}{\sin\f{a\pi}{k+2}}\right\}\nn\\
&=&\f{2}{k+2}\sum^{k+1}_{a=1}\sin^2\f{a\pi}{k+2}\prod_{r=1}^N\sum_{j_1,\cdots, j_N}\left\{\f{\sin\f{(2j_r+1)a\pi}{k+2}}{\sin\f{a\pi}{k+2}}\right\}\nn\\
&=&\f{2}{k+2}\sum^{k+1}_{a=1}\sin^2\f{a\pi}{k+2}\left[\sum_{j}\left\{\f{\sin\f{(2j+1)a\pi}{k+2}}{\sin\f{a\pi}{k+2}}\right\}\right]^N\nn
\ea
Using  Multinomial Expansion, the above expression can be recast into the following form
\ba
\Omega(N,k)&=&\f{2}{k+2}\sum^{k+1}_{a=1}\sin^2\f{a\pi}{k+2}\sum_{\left\{s_j\right\}}\f{N!}{\prod_js_j!}\prod_{j}\left\{\f{\sin\f{(2j+1)a\pi}{k+2}}{\sin\f{a\pi}{k+2}}\right\}^{s_j}\nn\\
&=&\sum_{\left\{s_j\right\}}\left[\f{2}{k+2}\sum^{k+1}_{a=1}\sin^2\f{a\pi}{k+2}\f{N!}{\prod_js_j!}\prod_{j}\left\{\f{\sin\f{(2j+1)a\pi}{k+2}}{\sin\f{a\pi}{k+2}}\right\}^{s_j}\right]\label{mul}
\ea
This is exactly eq.(\ref{nk}), which is nothing but eq.(\ref{ms}) written as sum over spin-configurations.

\section{`Quantum Hairs' and Microcanonical Ensemble}\label{microensemble}

 The next task is to define the microcanonical ensemble of QIHs by identifying and fixing the parameters of the theory which characterize the macrostates of a QIH. Right at the beginning, it has already been mentioned that $k$ and $N$ are the two independent integer-valued parameters, whose values, once specified, determine the number of microstates, for the corresponding macrostate of the QIH, given by the eq.(\ref{mul}). Thus, the role played by $k$ and $N$  is analogous to the role played by $M, J, Q$ in the classical theory where the classical black hole is characterized by these parameters called {\it hairs}. This is the rationale behind the use of the term `quantum hair' for both $k$ and $N$ which are {\it integer-valued} in the quantum theory and the word `hair' signifies the fact that they characterize the macrostates of a QIH, similar to the classical hairs $M, J, Q$ characterizing a black hole in the classical theory. To be precise, the term `quantum hair' was originally proposed in \cite{gp} to designate the parameter $N$ only. But, the terminology goes as well for $k$ also, as far as the quantum theory is concerned.

 To obtain the microcanonical entropy, one must maximize its expression with respect to the spin configurations, to determine the most probable spin configuration, subject to the restriction that the mean QIH area must equal the classical area up to $O(\lp^2)$ and the number $N$ of punctures is fixed. 

The area eigenvalue equation for a particular eigenstate of the QIH in the configuration basis can be written as
\ba
\hat A|\left\{s_j\right\}\rangle=8\pi\g\lp^2\sum_{j}s_j\j ~|\left\{s_j\right\}\rangle
\ea
Hence, the expectation value of the area operator for the QIH is given by
\ba
\langle\hat A\rangle=\langle\Psi_S|\hat A|\Psi_S\rangle&=&8\pi\g\lp^2\sum_{\left\{s_j\right\}}\omega[\left\{s_j\right\}]\sum_{j}s_j\j
~=~A_{cl} \pm O(\lp^2)
\ea
where $A_{cl}$ is the area of the classical IH closely represented by the QIH. Scaling the equation by $8\pi\g\lp^2$ and using $\langle\hat A\rangle/8\pi\g \lp^2\approx A_{cl}/8\pi\g \lp^2=k/2$ \cite{qg2,kras}, we obtain
\ba
\sum_{\left\{s_j\right\}}\omega[\left\{s_j\right\}]\sum_{j}s_j\j=\f{k}{2}
\ea
In this process we can also avoid the involvement of the ambiguous parameter $\g$ which will be ultimately fixed at the end by the usual argument of the validity of the BHAL.

Apart from this, the expectation value of the number of punctures for the QIH is given by 
\ba
\langle\hat N\rangle=\langle\Psi_S|\hat N|\Psi_S\rangle=\sum_{\left\{s_j\right\}}\omega[\left\{s_j\right\}]\sum_{j}s_j~=~N
\ea
where $\hat N$ can be considered as the operator for the number of punctures for a QIH. One should note that unlike the case of area the expectation value of $\hat N$ is {\it exactly} equal to the total number of punctures $N$.
Now, since we are doing equilibrium statistical mechanics (thus stability is assumed) of QIH with large area $(A_{cl}\gg\lp^2)$ and large number of punctures $(N\gg1)$, for all practical purposes, we can neglect the fluctuations and also consider that the  dominant contribution to the entropy comes from a most probable configuration, $\{\sj\}$ i.e. $\omega[\{\sj\}]\simeq 1$ \cite{pathria}.  Thus, every spin configuration $\{s_j\}$ must obey the following constraints
\begin{subeqnarray}\label{con}
&&{\cal C}_1 : \sum_{j} s_j = N \slabel{con1} \\
&&{\cal C}_2 : \sum_{j} s_j \sqrt{j(j+1)} = \f{k}{2}\slabel{con2} 
\end{subeqnarray}
of which $\{\sj\}$ will be the most probable one. Hence, we define a microcanonical ensemble of QIHs by assigning fixed values of $k$ and $N$ respectively. 
The obvious next step is the computation of the microcanonical entropy of a QIH whose {\it macrostates} are characterized by $k$ and $N$.

\section{Microcanonical Entropy}\label{microentropy}

Having defined the microcanonical ensemble appropriately, we shall now derive the microcanonical entropy of a QIH. The  microcanonical entropy of a QIH for given values of $k$ and $N$ is written as 
\ba
S_{MC}=\log\Omega(N,k)\simeq\log\Omega[\{\sj\}]
\ea
where we have set the Boltzmann constant to unity. Variation of $\log \Omega[\left\{s_j\right\}]$ with respect to $s_j$, subject to the constraints ${\cal C}_1$ and ${\cal C}_2$, yields the distribution function for the most probable configuration $\{\sj\}$ which maximizes the entropy of the QIH. In other words, $\sj$ satisfies the variational equation written as 
\ba
\d \log \Omega[\left\{s_j\right\}] -\s\sum_j\d s_j -\lm\sum_j\d s_j\j=0\label{var}
\ea 
where $\d$ represents variation with respect to $s_j$, $\s$ and $\lm$ are the Lagrange multipliers for ${\cal C}_1$ and ${\cal C}_2$ respectively. This yields the most probable distribution given by
\ba
\sj=N\exp\left[-\lm\sqrt{j(j+1)}-\s+\f{\d}{\d s_j}\log g[\{s_j\}]\right]\label{dc}		\ea

To proceed further we calculate $g[\{s_j\}]$ explicitly using saddle point approximation in the limit $k,N\to\infty$, which is appropriate for large black holes. First of all, we rewrite eq.(\ref{gsj}) replacing the summation over $a$ by integration as
\ba
g[\{s_j\}]&\simeq&\f{2}{k+2}\int^{k+1}_{1}\sin^2\f{a\pi}{k+2}\prod_{j}\left\{\f{\sin\f{a\pi(2j+1)}{k+2}}{\sin\f{a\pi}{k+2}}\right\}^{s_j} da\nn\\
&=&\f{2}{\pi}\int^{\pi -\e}_{\e}\sin^2\th ~\prod_{j}\left\{\f{\sin (2j+1)\th}{\sin \th}\right\}^{s_j} d\th
\ea
where we have applied a change in the integration variable as $a\pi/(k+2)=\th$ and for which the limits follow with $\e=\pi/(k+2)$. Now, for $k\to\infty$, $\e\to 0$. Hence, we can safely write
\ba
g[\{s_j\}]&\simeq&\f{2}{\pi}\int^{\pi}_{0}\sin^2\th ~\prod_{j}\left\{\f{\sin (2j+1)\th}{\sin \th}\right\}^{s_j} d\th\nn\\
&=&\f{1}{\pi}\int^{\pi}_{0}\exp\left[G(\th,k) \right]d\th - \f{1}{\pi}\int^{\pi}_{0}\exp\left[\ln(\cos 2\th)+G(\th,k) \right]d\th
\ea
where $G(\th,k)=\sum_{j}s_j\log\left\{\f{\sin (2j+1)\th}{\sin \th}\right\}$. The above two integrations can be performed by the saddle point method. To begin with, it is straightforward to show that
\begin{subeqnarray}\label{sp}
&&\lim_{\tn\to 0}G^{\prime}(\th,k)|_{\tn}=0\slabel{sp1}\\
&&\lim_{\tn\to 0}\{-2\tan 2\th+G^{\prime}(\th,k)\}|_{\tn}=0\slabel{sp2}
\end{subeqnarray}
which implies that the saddle point $\tn\simeq 0$ ($~~^{\prime}~$ denotes partial derivative with respect to $\th$). Now, Taylor expanding $G(\th,k)$ about the saddle point $\tn$ up to second order and applying the saddle point conditions from eqs.(\ref{sp}), we have
\ba
g[\{s_j\}]\simeq\f{1}{\pi}\prod_j(2j+1)^{s_j}\left[\int^{\pi}_{0}e^{-\f{1}{2}\alpha\xi^2}d\xi - \int^{\pi}_{0}e^{-\f{1}{2}(4+\alpha)\xi^2}d\xi\right]
\ea
where we have used $\xi=\th-\tn$ and the following limits
\ba
&&\lim_{\tn\to 0}\exp[G(\tn,k)]=\prod_j(2j+1)^{s_j}\nn\\
&&\lim_{\tn\to 0}G^{\prime\prime}(\th,k)|_{\tn}=-4\sum_j s_j~ j(j+1)\equiv -\alpha\nn\\
&&\lim_{\tn\to 0}\sec^22\tn=1\nn
\ea
Evaluation of the integral yields
\ba
g[\{s_j\}]\simeq\f{1}{\sqrt{2\pi}}\prod_j(2j+1)^{s_j}\left(\sqrt{\f{1}{\alpha}}~\text{Erf}\left[\pi\sqrt{\alpha/2}\right]-\sqrt{\f{1}{4+\alpha}}~\text{Erf}\left[\pi\sqrt{(4+\alpha)/2}\right]\right)\label{re}
\ea
The quantity $\alpha$ results from the second order approximation. Hence, while calculating $\alpha$ we can only use the results up to first order i.e. we use $\sj$ in place of $s_j$ whose expression will be given by  
\ba
\sj\simeq N (2j+1)\exp[-\lm\j -\s]\label{mpd1}
\ea
which follows from the fact that $g[\{s_j\}]\simeq \f{2C}{\pi}\prod_j(2j+1)^{s_j}$ neglecting the second order corrections, $C$ being some constant. Now, using eq.(\ref{mpd1}) in eq.(\ref{con1}) and eq.(\ref{con2}), one obtains 
\begin{subeqnarray}\label{cons}
\exp [{\s}] &=&\sum_{j=1/2}^{k/2} (2j+1)\exp[-\lm\j]\slabel{cons1}\\
k/2 &=&N\sum_{j=1/2}^{k/2} \j(2j+1)\exp[-\lm\j-\s]\slabel{cons2}
\end{subeqnarray}
\par
Using eq.(\ref{mpd1}), eq.(\ref{cons1}) and eq.(\ref{cons2}), it is straightforward to  show that $\alpha\simeq8N(d^2\s/d\lm^2)$.
Thus, $\alpha\to\infty$ for $N\to\infty$. Hence, we can approximately write $\text{Erf}\left[\pi\sqrt{(4+\alpha)/2}\right]\approx\text{Erf}\left[\pi\sqrt{\alpha/2}\right]$. Plotting the function $\text{Erf}\left[\pi\sqrt{\alpha/2}\right]$ with $\alpha$ one can see that the function attains a constant value for large $\alpha$. Hence, we can take $\lim_{\alpha\to\infty}\text{Erf}\left[\pi\sqrt{(4+\alpha)/2}\right]\simeq\lim_{\alpha\to\infty}\text{Erf}\left[\pi\sqrt{\alpha/2}\right]\simeq K$, some constant. Therefore, from eq.(\ref{re}) it follows that 
\ba
g[\{s_j\}]&\simeq&\f{K}{\sqrt{2\pi}}\prod_j(2j+1)^{s_j}\left(\sqrt{\f{1}{\alpha}}-\sqrt{\f{1}{4+\alpha}}\right)\nn\\
&\simeq&K\sqrt{\f{2}{\pi}}\prod_j(2j+1)^{s_j}\alpha^{-\f{3}{2}}\nn
\ea
Therefore, we have
\ba
\f{\d}{\d s_j}\log g[\{s_j\}]=\log (2j+1)-\f{6}{\alpha}j(j+1)\nn
\ea
Hence, considering variation of $\alpha$ resulting from the inclusion of the quadratic fluctuations, the distribution for the most probable configuration given by eq.(\ref{mpd1}) gets modified into
\ba
\sj\simeq N (2j+1)\exp[-\lm\j -\s-\f{6}{\alpha}j(j+1)]\label{mpd2}
\ea
Now, using the most probable distribution given by eq.(\ref{mpd1}) or eq.(\ref{mpd2}) (result will differ by a constant only) we calculate the microcanonical entropy of a QIH for given values of $k$ and $N$ using Stirling approximation and the result comes out to be
\ba
S_{MC}=\f{\lm k}{2}+N \s-\f{3}{2}\log N-\f{3}{2}\log (d^2\s/d\lm^2)+\cdots\label{smc1}
\ea
where $\lm$ and $\s$ satisfy the two equations (\ref{cons1}) and (\ref{cons2}). 

\par
As a consequence of what has been done up till now, one can show that the `quantum hair' $N$ is directly proportional to the classical quantity $A_{cl}$ at equilibrium. Using eq.(\ref{cons1}) and eq.(\ref{cons2}) and also finding $d\s/d\lm$ from eq.(\ref{cons1}) one can immediately show that 
\begin{subeqnarray}\label{estate}
k&=&-2N(d\s/d\lm)~~~~~~~~~~~~~~~~~~~~~~~~~~~~~~~~~~~~~~~~~~\slabel{estate1}\\
\text{or, using  $k=A_{cl}/4\pi\g\lp^2$ ,~~~ }\nn\\
A_{cl}&=&-8\pi\g\lp^2N(d\s/d\lm)\slabel{estate2}
\end{subeqnarray}   
 Eq.(\ref{estate1}) can be regarded as the `equation of state' for the QIH at equilibrium, whereas, eq.(\ref{estate2}) has the significance lying in the fact that there is indeed a deep underlying relationship between the so called `quantum hair' defined only in the quantum domain and the physically measurable quantity $A_{cl}$ which is purely classical. Hence it is justified why $N$ should play a very fundamental role in the thermodynamics of black holes if one tries to investigate by beginning from the underlying quantum theory provided by the QIH framework in LQG.

\par
Now, using eq.(\ref{estate2}) and the relation $k=A_{cl}/4\pi\g\lp^2$ the microcanonical entropy can be expressed  as
\ba
S_{MC}=\left[\f{f(\lm)}{2\pi\g}\right]\f{A_{cl}}{4\lp^2}-\f{3}{2}\log \f{A_{cl}}{4\lp^2}+\cdots\label{smc3}
\ea
where $f(\lm)=\lm-\left(\s/\f{d\s}{d\lm}\right)$. It is quite clear from the above expression for the {\mic} entropy that somehow we have to set the factor $f(\lm)/2\pi\g$ to be unity to obtain the Bekenstein-Hawking area law(BHAL). Let us see how can we do that.

In the limit $k\to\infty$, eq.(\ref{cons1}) and eq.(\ref{cons2})  can be approximated to be
\ba
e^{\s}&=&\f{2}{\lm^2}\left(1+\f{\3}{2}\lm\right)e^{-\f{\3}{~2}\lm}\label{rel1}\\
\f{k}{N}&=&1+\f{2}{\lm}+\f{4}{\lm(\3\lm+2)}\label{rel2}
\ea
Since we are dealing with the {\mic} ensemble, $k$ and $N$ are the given quantities which determine the Lagrange multipliers $\lm$ and $\s$ from the solutions of the above equations (\ref{rel1}) and (\ref{rel2}). Hence, in the {\mic} ensemble, $\lm\equiv\lm(k/N)$ and $\s\equiv\s(k/N)$ are functions of $k$ and $N$.  It should be noted that {\it there is no freedom to choose $\lm$ (hence $f(\lm)$) or $\s$ in the {\mic} ensemble.} Strictly speaking, we should write the function $f$ as $f(k/N)$ and corresponding to every value of $k/N$ there exists a unique value of $f$. To retrieve the BHAL from eq.(\ref{smc3}) we can choose $\g=f(k/N)/2\pi$, 
which results in
\ba
S_{MC}(A_{cl})=\f{A_{cl}}{4\lp^2}-\f{3}{2}\log \f{A_{cl}}{4\lp^2}+\cdots\label{smc4}
\ea
Hence, the range of allowed values of $\g$ will be dictated by the range of $f$. Now, one can find $f(\lm)$ to be given by the following expression
\ba
f(\lm)=\lm\left[1+\f{2(2+\3\lm)\{\log 2 - 2\log \lm+\log(1+\3\lm/2)-\3\lm/2\}}{\3\lm^2+2(\3+1)\lm+8}\right]
\ea
From eq.(\ref{rel2}) it is evident that for any value of $\lm$ between $0$ and $\infty$, $k/N$ remains positive and from the above expression it is evident that for $0<\lm<\infty$, we have $0<f(\lm)<\infty$. Hence, without finding the explicit form of $f(k/N)$ one can remain assured that $f$ takes values from $0$ to $\infty$ for positive values of $k/N$. It follows that $\g$ can take value from $0$ to $\infty$ for the leading term of the {\mic} entropy to follow the BHAL.

{\it Remarks :} One can look into \cite{majhi1} where a similar argument about fixing $\g$ has been given remaining within the {\mic} ensemble. The difference between the result of \cite{majhi1} and the present work is in the range of allowed values of $\g$. But this is not a contradiction. The crucial point is how we choose $\g$ to obtain our desired form of the {\mic} entropy. In this work we choose $\g$ so as to obtain the BHAL as the leading term, whereas, in \cite{majhi1}, $\g$ is chosen to obtain the form of the {\mic} entropy derived in \cite{gp} where an extra term $N\s$ should br present. Since, in LQG calculations of {\mic} entropy from kinematical Hilbert space, $\g$ is determined by demanding the form of the entropy we want, in this scenario where the total number of punctures $N$ is considered as a macroscopic parameter besides $k$, there are two ways we can fix $\g$. One of which has been shown here and the other is what has been done in \cite{majhi1}. There is no contradiction between these two works.

\section{Discussion}\label{disc}

As mentioned in the Introduction, the LQG formulation of a QIH in terms of a CS theory coupled to spins directly involves the use of the integer parameters (`quantum hairs') $k$ and $N$. Strictly speaking, if the classical area $A_{cl}$ is taken to be a hair, the BI parameter $\gamma$ (the coefficient of a topological contribution to the classical action from the Nieh-Yan invariant \cite{date}) is considered as an independent coupling parameter, then the only new parameter that has appeared in the quantum theory is the number of punctures $N$. Equivalently, one can, as we have in this paper, take $k$ and $N$ to be the parameters characterizing the quantum theory. It is significant that the requirement that the microcanonical entropy of a QIH yields the BHAL for large $k~,~N$ does not only allow the BI parameter $\gamma$ to take any positive value on the real line, it also yields the subleading logarithmic correction derived in earlier literature with the universal coefficient $-3/2$. Thus, the complete characterization of the {\it macrostate} of a QIH is given by two independent parameters, namely, $k$ and $N$.  The limit of large $k$ and $N$ can thus be taken to be the approximately the semiclassical domain, since it is in this limit that the computation reliably yields an answer for the microcanonical entropy which can be explicitly expressed entirely in terms of the Bekenstein-Hawking entropy. Although the `classical limit' of bulk LQG has certain ambiguities in extraction of a classical metric from expectation values of geometrical observables within coherent states, as far as classical horizons are concerned, such an ambiguity can perhaps be avoided by working with the idea of an effective QIH in this limit and comparing with semiclassical behaviour. Of course, our approach does not take into account the entanglement of quantum matter states in the bulk and the boundary, not entanglement between quantum matter and quantum spacetime states, but accounts only for entanglement between bulk and boundary quantum spacetime states. The inclusion of quantum matter together with quantum geometry thus remains one of the key points of our research agenda for the immediate future.   

The departure from recent literature is the direct link established in this paper of these results and interpretation within the CS formulation of a QIH. The point here is that the computation of the microcanonical entropy of a QIH is {\it not} merely a combinatoric exercise in statistical mechanics of an ideal gas of punctures with some restrictions gleaned out of LQG as in some part of the recent literature \cite{gp,gm1,gm2,gm3,gm4,dole,mei}, with at best a weak link to the complete formulation as a CS theory that has been available for years. Here we have shown that the original formulation yields an understanding that is complete as a quantum theory. 

An extension of the foregoing analysis to Trapping horizons and Entanglement entropy of matter and radiation fields in their vicinity, would amount to truly new physics of radiant quantum horizons beyond general relativity. This has within it the potential to surpass, because of its firmer quantum geometric underpinning, recently proposed speculative ideas of a somewhat ad hoc nature based on semiclassical analysis \cite{mathur,sam,marpol}, about how classical horizon geometry must change (become a `fuzzball' or an `energetic curtain' or a `firewall') so as to allow the existence of well-defined scattering amplitudes for quantum matter fields. Whether or not such a structure emerges from LQG in an appropriate limit is not known at the moment, since there is still no complete quantum geometric analysis of Hawking radiation and its various conundra from an LQG standpoint. The problem at hand involves a quantization of the kinematical phase space of a Trapping/Dynamical horizon along the lines of \cite{qg1,qg2} which may not be technically so simple as a QIH because of the transient behaviour inherent in the geometry. However, perhaps a perturbation of the CS description by a set of appropriately chosen matter field operators might serve as a first approximation to the problem. The strength of the LQG approach lies in the transparent manner in which the CS symplectic structure emerges from the boundary conditions in the incipient formulation \cite{qg2}, without having to rely on conjectured results. The relation of the CS Hilbert space to the conformal blocks of the WZW model on the QIH is also not a matter of conjecture for large black holes. Thus, the LQG analysis of a QIH geometry throws up holographic structures as {\it emergent}, without any prior notion that they have to be there. One expects that generalizations to Trapping horizons to centre around this theme so as to reap the benefits of the 4 dimensional gravity - 2 dimensional conformal field theory relation found in ref. \cite{km98}. Note however that the thermal stability of radiant trapping horizons which approach an equilibrium QIH has been discussed within the LQG framework yielding a criterion of stability involving the equilibrium mass and the microcanonical entropy of the QIH in ref. \cite{stab1,stab2,stab3}.

\par
{\bf Acknowledgment :} We thank Marc Geiller for a careful reading of an earlier version of this manuscript and for offering some incisive comments which have helped us to improve some parts of this article. We also thank an anonymous referee for pointing out a few typo in an earlier version of this paper. One of us, A. M.  sincerely acknowledges the financial support provided by the Department of Atomic Energy, India for pursuing his research goals.


\begin{thebibliography}{777}

\bibitem{qg1} A. Ashtekar, J. Baez, A. Corichi and K. Krasnov, {\it Phys. Rev. Lett.} {\bf 80} (1998) 904-907; \href{http://arxiv.org/abs/gr-qc/9710007v1}{arXiv:gr-qc/9710007v1}

\bibitem{qg2} A. Ashtekar, J. Baez and K. Krasnov, {\it Adv. Theor. Math. Phys.} {\bf 4} (2000) 1; \href{http://arxiv.org/abs/gr-qc/0005126}{arXiv:gr-qc/0005126v1}

\bibitem{abf} A. Ashtekar, A. Corichi and K. Krasnov, {\it Adv. Theor. Math. Phys.} {\bf 3} (2000) 419-478; \href{http://arxiv.org/abs/gr--qc/9905089}{arXiv:gr-qc/9905089}

\bibitem {km11} R. K. Kaul and P. Majumdar, {\it Phys. Rev.} {\bf D83} (2011) 024038; \href{http://arxiv.org/abs/1004.5487}{arXiv:1004.5487}

\bibitem{sol} S. N. Solodukhin, {\it Phys. Rev.} {\bf D51} (1995) 618-621; \href{http://arxiv.org/abs/hep-th/9408068}{arXiv:hep-th/9408068v1}

\bibitem{sol1} R. B. Mann and S. N. Solodukhin , 	{\it Nucl. Phys.} {\bf B523} (1998) 293-307; \href{http://arxiv.org/abs/hep-th/9709064}{arXiv:hep-th/9709064v4}

\bibitem{sol2} S. N. Solodukhin, {\it Phys. Rev.} {\bf D51} (1995) 609-617; \href{http://arxiv.org/abs/hep-th/9407001}{arXiv:hep-th/9407001v1}

\bibitem{fursaev} D. V. Fursaev, {\it Phys. Rev.} {\bf D51} (1995) 5352-5355; \href{http://arxiv.org/abs/hep-th/9412161}{arXiv:hep-th/9412161v1}
 
\bibitem{sen1} A. Sen, {\it Logarithmic Corrections to Schwarzschild and Other Non-extremal Black Hole Entropy in Different Dimensions}; \href{http://arxiv.org/abs/1205.0971}{arXiv:1205.0971v1}

\bibitem{sen2} S. Banerjee, R. K. Gupta, I. Mandal and A. Sen, {\it JHEP} {\bf 1111} (2011) 143 ; \href{http://arxiv.org/abs/1106.0080}{arXiv:1106.0080v2}

\bibitem{sen3} A. Sen, {\it Gen. Rel. Grav.} {\bf 44} (2012) 1207-1266; \href{http://arxiv.org/abs/1108.3842}{arXiv:1108.3842v1}


\bibitem{sen4} A. Sen, {\it Gen. Rel. Grav.} {\bf 44} (2012) 1947-1991; \href{http://arxiv.org/abs/1109.3706}{arXiv:1109.3706v3}


\bibitem{bomb} L. Bombelli, R. K. Koul, J. Lee and R. D. Sorkin, {\it Phys. Rev.} {\bf D34} (1986) 373.

\bibitem{sred} M. Srednicki, {\it Phys. Rev. Lett.} {\bf 71} (1993) 666.

\bibitem{hay} S. A. Hayward, {\it Dynamics of black holes}; \href{http://arxiv.org/abs/0810.0923}{arXiv:0810.0923v2} and references cited therein

\bibitem{ashkris} A. Ashtekar and B. Krishnan, {\it Living Rev. Relativity} {\bf 7} (2004) 10;\href{http://arxiv.org/abs/gr-qc/0407042}{arXiv:gr-qc/0407042v3}

\bibitem{pranz} D. Pranzetti, {\it Phys. Rev. Lett.} {\bf 109}, (2012) 011301 , \href{http://arxiv.org/abs/1204.0702}{arXiv:1204.0702v1} ; {\it Dynamical evaporation of quantum horizons}, \href{http://arxiv.org/abs/1211.2702}{arXiv:1211.2702v1} ;  see also \cite{kras} for an intuitive picture of Hawking radiation  from a quantum viewpoint which was proposed long before the advent of dynamical horizons. Classical approaches are made in \cite{ccg} and \cite{hay1}.

\bibitem{kras} K. V. Krasnov, {\it Gen. Rel. Grav.} {\bf 30} (1998) 53-68; \href{http://arxiv.org/abs/gr-qc/9605047}{arXiv:gr-qc/9605047v3}

\bibitem{ccg} A. Chatterjee, B. Chatterjee and A. Ghosh, {\it Hawking radiation from dynamical horizons}, \href{http://arxiv.org/abs/1204.1530}{arXiv:1204.1530v1}

\bibitem{hay1} S. A. Hayward, R. D. Criscienzo, M. Nadalini, L. Vanzo and S. Zerbini, {\it AIP Conf. Proc.} {\bf 1122}:145-151 (2009), \href{http://arxiv.org/abs/0812.2534}{arXiv:0812.2534v1}; {\it Local Hawking temperature for dynamical black holes}, \href{http://arxiv.org/abs/0806.0014}{arXiv:0806.0014v2}


\bibitem{gp} A.Ghosh and A.Perez, {\it Phys. Rev. Lett.} {\bf 107}, (2011) 241301; \href{http://arxiv.org/abs/1107.1320}{arXiv:1107.1320v2}

\bibitem{km98} R. K. Kaul and P. Majumdar, {\it Phys.Lett.} {\bf B439} (1998) 267-270; \href{http://arxiv.org/abs/gr-qc/9801080}{arXiv:gr-qc/9801080v2}

\bibitem{pathria} L. D. Landau and E. M. Lifschitz, {\it Statistical Physics}, Pergamon Press, 1980; D' ter Haar, {\it Elements of Statistical Mechanics}, 3rd Edition, Butterworth-Heinemann, 1995;
R. K. Pathria, P. D. Beale, {\it Statistical Mechanics}, 3rd Edition, Elsevier, 2011

\bibitem{majhi1} A. Majhi, {\it Class. Quant. Grav.}  {\bf 31} (2014) 095002, \href{http://arxiv.org/abs/1205.3487}{arXiv:1205.3487v2}


\bibitem{km00} R. K. Kaul and P. Majumdar, {\it Phys. Rev. Lett.} {\bf 84} (2000) 5255-5257; \href{http://arxiv.org/abs/gr-qc/0002040}{arXiv:gr-qc/0002040v3} 

\bibitem{per1} J. Engle, K. Noui and A. Perez, {\it Phys. Rev. Lett.} {\bf 105} (2010) 031302;\href{http://arxiv.org/abs/0905.3168}{arXiv:0905.3168}

\bibitem{per2} J. Engle, K. Noui, A. Perez and D. Pranzetti, {\it Phys.Rev.} {\bf D82} (2010) 044050; \href{http://arxiv.org/abs/1006.0634}{	arXiv:1006.0634v1}

\bibitem{per3} J. Engle, K. Noui, A. Perez and D. Pranzetti, {\it JHEP} {\bf 1105} (2011) 016; \href{http://arxiv.org/abs/1103.2723v1}{arXiv:1103.2723v1 }


\bibitem{per4} A. Perez and D. Pranzetti, {\it Entropy} {\bf 13} (2011) 744-777; \href{http://arxiv.org/abs/1011.2961v3}{arXiv:1011.2961v3}

\bibitem{dole}M. Domagala and J. Lewandowski, {\it Class. Quant. Grav.} {\bf 21} (2004) 5233-5244; \href{http://arxiv.org/abs/gr-qc/0407051}{arXiv:gr-qc/0407051v2}

\bibitem{mei}K. A. Meissner, {\it Class. Quant. Grav.} {\bf 21} (2004) 5245-5252; \href{http://arxiv.org/abs/gr-qc/0407052v1}{arXiv:gr-qc/0407052v1}

\bibitem{kaul12} R. K. Kaul, {\it SIGMA} {\bf 8, 005} (2012); \href{http://arxiv.org/abs/1201.6102}{arXiv:1201.6102v2} 

\bibitem{ashlew} A. Ashtekar and J. Lewandowski, {\it Class. Quant. Grav.} {\bf 21} (2004) R53, \href{http://arxiv.org/abs/gr-qc/0404018v2}{arXiv:gr-qc/0404018v2}


\bibitem{wit} E. Witten, {\it Commun. Math. Phys} {\bf 121} (1989) 351-399


\bibitem{date} G. Date, R. K. Kaul and S. Sengupta, {\it Phys. Rev.} {\bf D79} (2009) 044008 \href{http://arxiv.org/abs/0811.4496}{arXiv:0811.4496v2}


\bibitem{gm1} A. Ghosh and P. Mitra, {\it Phys. Rev.} {\bf D71} (2005) 027502; \href{http://arxiv.org/abs/gr-qc/0401070}{arXiv:gr-qc/0401070v3}


\bibitem{gm2} A. Ghosh and P.Mitra, {\it Phys. Lett.} {\bf B616} (2005) 114-117; \href{http://arxiv.org/abs/gr-qc/0411035}{arXiv:gr-qc/0411035v3}


\bibitem{gm3} A. Ghosh and P. Mitra, {\it Phys. Rev.} {\bf D74} (2006) 064026; \href{http://arxiv.org/abs/hep-th/0605125}{arXiv:hep-th/0605125v2}


\bibitem{gm4} A. Ghosh and P.Mitra, \textit{Phys. Rev. Lett.} {\bf 102} 141302 (2009); \href{http://arxiv.org/abs/0809.4170v3}{arXiv:0809.4170v3}

\bibitem{mathur} S. D. Mathur, {\it JHEP} {\bf 1007} (2010) 009, {\it Erratum-ibid} {\bf 1104} (2011) 032 ; \href{http://arxiv.org/abs/1004.4142}{arXiv:1004.4142v2} and references cited therein


\bibitem{sam} S. L. Braunstein, S. Pirandola and K. Życzkowski, {\it Phys. Rev. Lett.} {\bf 110}, 101301 (2013), \href{arxiv.org/abs/0907.1190}{arXiv:0907.1190v3}

\bibitem{marpol} A. Almheiri, D. Marolf, J. Polchinski and J. Sully, {\it Black Holes: Complementarity or Firewalls?}, \href{http://arxiv.org/abs/1207.3123}{arXiv:1207.3123v3} and references cited therein


\bibitem{stab1} A. Chatterjee and P. Majumdar, {\it Phys. Rev.} {\bf D72} (2005) 044005; \href{http://arxiv.org/abs/gr-qc/0504064}{arXiv: gr-qc/0504064}


\bibitem{stab2} P. Majumdar, {\it Class. Quant. Grav.} {\bf 24} (2007) 1747; \href{http://arxiv.org/abs/gr-qc/0701014}{arXiv: gr-qc/0701014}


\bibitem{stab3} A. Majhi and P. Majumdar,  {\it Class. Quant. Grav.} {\bf 29} (2012) 135013; \href{http://arxiv.org/abs/1108.4670}{arXiv:1108.4670v1}


\end{thebibliography}
\end{document}